\documentclass{PoS}

\usepackage{epsf}
\usepackage{amssymb}
\usepackage{amsmath}
\usepackage{amsfonts}
\usepackage{cite}
\usepackage[small]{caption2}

\newcommand{\zbar}{\bar{z}}
\newcommand{\odd}{\mathbb{O}}
\newcommand{\pom}{\mathbb{P}}

\newcommand{\gev}{{\rm GeV}}

\usepackage[T1]{fontenc}
\usepackage{psfrag,epsfig,graphicx,graphics}

\newcommand{\be}{\begin{equation}}
\newcommand{\beq}{\begin{equation}}
\newcommand{\ee}{\end{equation}}
\newcommand{\eq}{\end{equation}}
\newcommand{\eeq}{\end{equation}}

\newcommand{\bea}{\begin{eqnarray}}
\newcommand{\eea}{\end{eqnarray}}

\def\slashchar#1{\setbox0=\hbox{$#1$}
   \dimen0=\wd0
   \setbox1=\hbox{/} \dimen1=\wd1
   \ifdim\dimen0>\dimen1
      \rlap{\hbox to \dimen0{\hfil/\hfil}}
      #1
   \else
      \rlap{\hbox to \dimen1{\hfil$#1$\hfil}}
      /gdatdafinal2.tex
   \fi}

\title{Pomeron Odderon interference in production of two $\pi^+-\pi^-$ pairs at LHC and ILC}

\ShortTitle{Pomeron Odderon interference in production of two $\pi-\pi$ pairs at LHC and ILC}

\author{B.~Pire\\
CPHT, {\'E}cole Polytechnique, CNRS, 91128 Palaiseau Cedex, France\\
E-mail: \email{pire@cpht.polytechnique.fr}}

\author{F.~Schwennsen\\
CPHT, {\'E}cole Polytechnique, CNRS, 91128 Palaiseau Cedex, France \ {\em \&}\\
 LPT, Universit{\'e} Paris-Sud, CNRS, 91405 Orsay, France\\
E-mail: \email{fschwenn@cpht.polytechnique.fr}}

\author{L.~Szymanowski\\
Soltan Institute for Nuclear Studies, PL-00-681 Warsaw, Poland\\
E-mail: \email{Lech.Szymanowski@fuw.edu.pl}}

\author{\speaker{S.~Wallon}
\\
        LPT, Universit{\'e} Paris-Sud, CNRS, 91405 Orsay, France \ {\em \&}\\
UPMC Univ. Paris 06, facult\'e de physique, 4 place Jussieu, 75252 Paris Cedex 05, France\\
        E-mail: \email{wallon@th.u-psud.fr}}

\abstract{We propose to look for the Odderon through the production of two pion pairs in photon collisions at high energies. We calculate the corresponding matrix elements in $k_T$-factorization and discuss the possibility to reveal the existence of the perturbative Odderon by charge asymmetries, relying on models for the generalized distribution amplitudes
of $\pi^+\pi^-$.
The application of this strategy to ultraperipheral collisions at the LHC
suffers from the difficulty to trigger on interesting events and is plagued
with severe background problems in $p$-$p$ mode. Electron - positron colliders like ILC  seem to better suit  this physics.
}

\FullConference{European Physical Society Europhysics Conference on High Energy Physics,
EPS-HEP 2009,\\
		 July 16 - 22 2009\\
		 Krakow, Poland}

\begin{document}


At high energies, amplitudes of reactions with rapidity gaps in hadronic interactions are dominated by the exchange of a color singlet, $C$-even state  -- called the Pomeron. In  perturbative QCD,
 the Pomeron can be described at lowest order as the exchange of two gluons in the color singlet state. 
Its $C$-odd partner, the Odderon, while  needed~\cite{LN}, has never been seen  in the perturbative regime, where  it can be described (at lowest order) by the exchange of three gluons in a color singlet state.
Due to its small exchange amplitude one should rather consider observables sensitive to interference effects.  We present here our results \cite{Pire:2008xe} on such an observable in the hard regime, a charge asymmetry  in the  production of two $\pi^+ \pi^-$ pairs in photon-photon collisions.


\begin{figure}[t]
\centerline{\includegraphics[height=4.0cm]{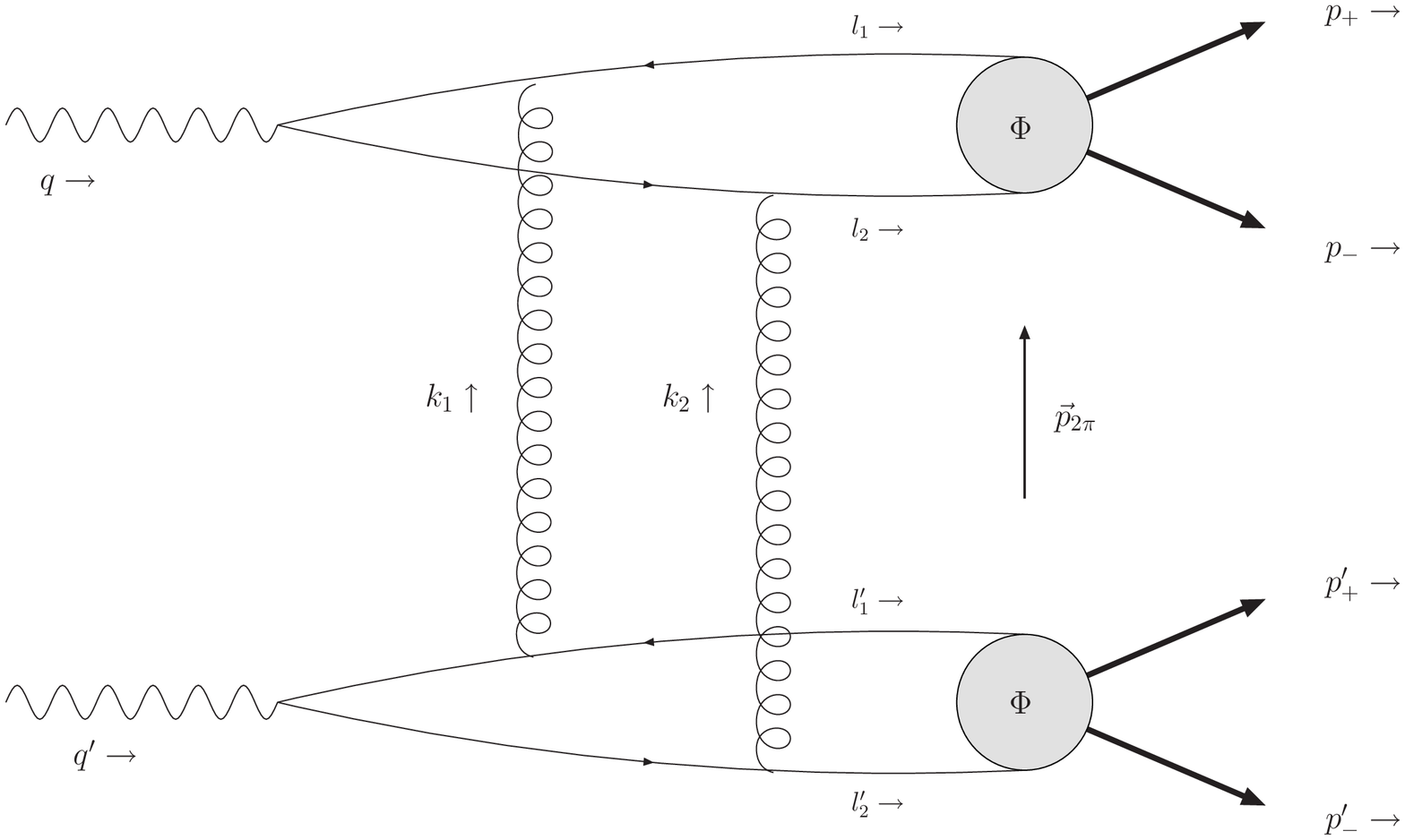}}
\caption{{\protect\small Kinematics of the reaction $\gamma \gamma \to \pi^+ \pi^-\;\; \pi^+ \pi^-$ in the two gluon exchange process.}}
\label{fig:1}
\end{figure}

Figure~\ref{fig:1} shows a sample diagram of the process under consideration. We consider large $\gamma\gamma$ energies such that the amplitude can be expressed in terms of two impact factors convoluted over the transverse momenta of the exchanged gluons. The impact factors are universal and consist of a perturbative part -- describing the transition of a photon into a quark-antiquark pair -- and a non-perturbative part, the two pion generalized distribution amplitude (GDA) parametrizing the quark-antiquark hadronization into the 
 pion pair~\cite{Polyakov:1998ze,Diehl:2000uv}.   This comes as a variant of the approach which has been previously proposed in the case of the electroproduction of a pion pair~\cite{Hagler:2002nh,Warkentin:2007su}, and which is based on the fact that  the $\pi^+ \pi^- $-state does not have any definite charge parity. These GDAs
 which are functions of the longitudinal momentum fraction $z$ of the quark,  of
the angle $\theta$ (in the rest frame of the pion pair) and of the invariant mass $m_{2\pi}$ of the pion pair are the only phenomenological inputs. A useful parametrization in terms of  Legendre polynomials $P_l(\beta\cos\theta)\,,$ where $\beta=\sqrt{1-4m_\pi^2/m_{2\pi}^2\,}$\,,
reads \cite{Polyakov:1998ze}:
\begin{eqnarray*}
  \Phi^{I=1} (z,\theta,m_{2\pi}) &=& 6z\zbar\beta f_1(m_{2\pi}) \cos\theta ,\\
  \Phi^{I=0} (z,\theta,m_{2\pi}) &=& 5z\zbar(z-\zbar)\left[-\frac{3-\beta^2}{2}f_0(m_{2\pi})+\beta^2f_2(m_{2\pi})P_2((\beta\cos\theta)\right],
\end{eqnarray*}
where $f_1(m_{2\pi})$ can be identified with the electromagnetic pion form factor $F_\pi(m_{2\pi})$. 
For the $I=0$ component we use~\cite{Pire:2008xe} different models partially following Ref.~\cite{Hagler:2002nh} and Ref.~\cite{Warkentin:2007su}.

\begin{figure}
  \centering
  \includegraphics[width=5cm]{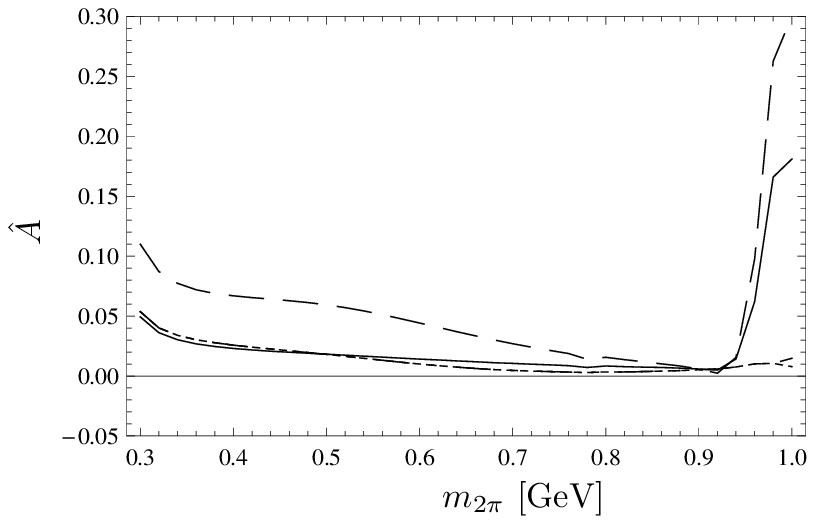}\hspace{10mm}
  \includegraphics[width=5cm]{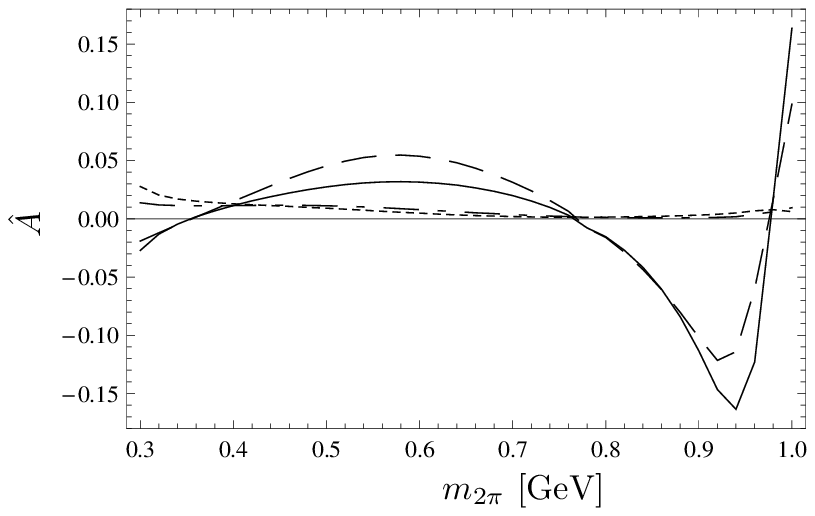}
  \caption{Asymmetry $\hat{A}$ at $t=-1\,\gev^2$ for various models of the GDAs. Left column has $m_{\rm min}=.3\,\gev$ and $m_{\rm max}=m_\rho$, while right column has $m_{\rm min}=m_\rho$ and $m_{\rm max}=1\,\gev$. 
}
\label{fig:asymplot1}
\end{figure}

The key to obtain an observable which linearly depends on the Odderon amplitude ${\cal M}_\odd$ is the orthogonality of the $C$-even GDA (entering ${\cal M}_\odd$) and the $C$-odd one (entering the Pomeron amplitude ${\cal M}_\pom$) in the space of  Legendre polynomials. We thus define the charge asymmetry as:
\begin{gather*}
 \hat{A}(t,m_{2\pi}^2;m_{\rm min}^2,m_{\rm max}^2) = \frac{\int_{m_{\rm min}^2}^{m_{\rm max}^2} dm_{2\pi}'^2\int\cos\theta\,\cos\theta'\,d\sigma(t,m_{2\pi}^2,m_{2\pi}'^2,\theta,\theta')}{\int_{m_{\rm min}^2}^{m_{\rm max}^2} dm_{2\pi}'^2\int\,d\sigma(t,m_{2\pi}^2,m_{2\pi}'^2,\theta,\theta')}   
\label{eq:ahat}
\end{gather*}
\begin{gather*}
= \frac{\int_{m_{\rm min}^2}^{m_{\rm max}^2} dm_{2\pi}'^2\int_{-1}^1d\cos\theta\int_{-1}^1d\cos\theta'\;2\cos\theta\,\cos\theta'\,{\rm Re}\left[\mathcal{M}_\pom(\mathcal{M}_\odd+\mathcal{M}_{\gamma})^*\right]}{\int_{m_{\rm min}^2}^{m_{\rm max}^2} dm_{2\pi}'^2\int_{-1}^1d\cos\theta\int_{-1}^1d\cos\theta'\,\left[\left|\mathcal{M}_\pom\right|^2+\left|\mathcal{M}_\odd+\mathcal{M}_{\gamma}\right|^2\right]}
. 
\end{gather*}
The result for $\hat{A}$ is shown in Fig.~\ref{fig:asymplot1} with two different choices for the integration regions. We stress that our framework is only justified for $m_{2\pi} ^2  \ll -t$.
Our Born order estimate should be corrected by BFKL effects which can be estimated semi-phenomenologically from HERA data (intercept $\alpha_\pom \simeq 1.3$), leading for $\sqrt{s}_{min} =20$ GeV to an increase of $\sim 6$ for the counting rates and to a slight decrease of the asymmetry $\sim {\cal M}_\odd /{\cal M}_\pom\,.$

The question naturally arises whether it is possible to measure this asymmetry in ultraperipheral collisions at the LHC, either in proton-proton or in ion-ion collisions.
Various difficulties in $p$-$p$ mode coming either from
difficulties to trigger on exclusive events, or from background separation,
seem to lead to a negative answer. In ion-ion collisions, the trigger problem may be solved by  detecting neutrons from giant dipole resonances in the Zero Degree Calorimeters, but
the rates are lower.
In contrast, an electron-positron collider such as the projected ILC would be the ideal environment to study the process under consideration. Photon-photon collisions are indeed the dominant processes there and no pile up phenomenon can blur the picture of a scattering event.
Studies of similar exclusive processes~\cite{2rho} teach us that high rates may be expected.

This work is supported in part by the Polish Grant N202 249235, the French-Polish scientific agreement Polonium, by the grant ANR-06-JCJC-0084 and by the ECO-NET program, contract 12584QK.



\begin{footnotesize}


\end{footnotesize}


\end{document}